\documentclass[prr,aps,twocolumn,longbibliography]{revtex4-2}
\usepackage{color}
\usepackage{graphicx}
\usepackage{bm}
\usepackage{amsmath}
\usepackage{amssymb}
\usepackage{amsthm}
\usepackage{amsfonts}
\usepackage{enumerate}
\usepackage{latexsym}
\usepackage{lipsum}
\usepackage[colorlinks=true,urlcolor=blue,citecolor=blue,allcolors=blue]{hyperref}

\usepackage{braket}
\usepackage{array}
\usepackage{float}
\usepackage{cleveref}

\begin{document}

\title{Bounding multifractality by observables}

\author{Tuomas I. Vanhala, Niklas J\"arvelin and  Teemu Ojanen}
\affiliation{Computational Physics Laboratory, Physics Unit, Faculty of Engineering and
Natural Sciences, Tampere University, P.O. Box 692, FI-33014 Tampere, Finland}
\affiliation{Helsinki Institute of Physics P.O. Box 64, FI-00014, Finland}

\begin{abstract}
  
  Fractal dimensions have been used as a quantitative measure for
  structure of eigenstates of quantum many-body systems, useful for
  comparison to random matrix theory predictions or to distinguish
  many-body localized systems from chaotic ones. For chaotic systems
  at midspectrum the states are expected to be ``ergodic'', infinite
  temperature states with all fractal dimensions approaching 1 in the
  thermodynamic limit. However, when moving away from midspectrum, the
  states develop structure, as they are expected to follow the
  eigenstate thermalization hypothesis, with few-body observables
  predicted by a finite-temperature ensemble. We discuss how this
  structure of the observables can be used to bound the fractal
  dimensions from above, thus explaining their typical arc-shape over
  the energy spectrum. We then consider how such upper bounds act as a
  proxy for the fractal dimension over the many-body localization
  transition, thus formally connecting the single-particle and Fock
  space pictures discussed in the literature.

\end{abstract}

\maketitle

\section{Introduction}

Thermalization, chaos, and ergodicity in isolated many-body quantum
systems have been frequent topics in recent research
\cite{Polkovnikov2016ETH,Deutsch_2018}. Each of these concepts is an
important part of the statistical mechanics of classical systems, and
each of them has inspired ways to characterize the eigenstates of
many-body quantum systems. The eigenstate thermalization hypothesis
(ETH) explains dynamical thermalization by transferring the thermal
properties to the eigenstates of the system
\cite{Polkovnikov2016ETH,Deutsch_2018,PhysRevE.50.888}, thus giving a
characterization of few-body observables. In a chaotic system, away
from spectral edges, such observables are expected to approach their
thermal expectation values that only depend on the energy of the
eigenstate. The observables reveal a structure in the state vector
that only vanishes at midspectrum, where the temperature approaches
infinity \cite{PhysRevE.107.024102} and observables carry no
information about the underlying Hamiltonian. This can be contrasted
to many-body localized states, where multiple ``emergent'' integrals
of motion are needed to describe the eigenstates
\cite{2017AnP...52900278I}, and structure is present even at
midspectrum. Indeed, it is possible to characterize the many-body
localization transition by defining a suitable measure for this
structure using e.g. one-particle occupation numbers
\cite{PhysRevLett.115.046603,2017AnP...52900356B,PhysRevB.103.214206,PhysRevB.97.104406,SciPostPhys.4.1.002,SciPostPhys.6.4.050,PhysRevA.101.063617}.

Another way to quantify structure of eigenstates is to consider the
full state vector in a fixed basis. This is conceptually related to
studies of quantum ergodicity, where highly excited eigenstates of
classically chaotic systems are found to have their weight evenly
distributed throughout the phase space \cite{Pilatowsky_Cameo_2024,PhysRevLett.53.1515,M_V_Berry_1977}.
In many-body systems, the space-filling properties of midspectrum
eigenstates have been compared to eigenstates of random matrix
ensembles with symmetry classes appropriate for the underlying
Hamiltonian \cite{Polkovnikov2016ETH}. Common quantitative measures
include the inverse participation ratio, and, more generally,
Renyi-$q$ entropies $S_q$ of the weight distribution of the state,
which have been studied for numerous many-body models
\cite{Polkovnikov2016ETH,Borgonovi_2016,PhysRevB.103.214206,PhysRevLett.123.180601,PhysRevE.98.022204,De_Luca_2013,Santos_2010,Beugeling_2015,PhysRevB.91.081103,Misguich_2016,Tsukerman_2017,PhysRevLett.124.200602}.
Denoting the Hilbert space size as $Q$, the complexity of state
$\ket{\psi}$ can be expanded as
\begin{equation}
  S_q(\ket{\psi})=D_q \log(Q) + O(1/Q),
\end{equation}
where $D_q$ is the fractal dimension in the thermodynamic
limit. Numerical results indicate that midspectrum states of chaotic
systems generally have $D_q=1$
\cite{PhysRevE.100.032117,PhysRevLett.123.180601,Luitz_2015}, meaning
that the state fills at least a finite fraction of the available
Hilbert space in the thermodynamic limit, in agreement with the
expectation that these are infinite temperature states with no visible
structure in few-body observables. However, as soon as we move away
from the infinite temperature point, structure appears and $0<D_q<1$
even in a chaotic system. The coefficient $D_q$ attains an arc shape
as a function of the energy, decreasing towards the edges of the
spectrum
\cite{Polkovnikov2016ETH,Santos_2010,PhysRevE.100.032117,Rigol_2010,PhysRevE.105.014109}.
Viewed like this, the state thus becomes ``fractal'', with an
effective ``volume'' scaling nontrivially slower than the volume of
the space. Similarly, the many-body localized states have $D_q<1$ even
at midspectrum, and numerical scaling analysis indicates that the
localization transition may be associated with a jump in the value of
$D_q$ in the thermodynamic limit \cite{PhysRevLett.123.180601}. In
either of these cases it is not surprising to find $D_q<1$, as
few-body observables already indicate that the state develops
structure, but the exact relation between these two points of view has
not fully been explored. Is it possible to predict $D_q$ if few-body
observables are assumed to be known?

In this work we discuss how few-body observables can be used to bound
the complexity $S_q$, and thus the fractal dimension $D_q$, of
eigenstates of many-body systems. We define an entropy-like quantity
closely related to the occupation number entropy discussed in
\cite{PhysRevLett.115.046603}, and show that it bounds the complexity
from above, thus supplementing the lower bound relation discussed in
our earlier work \cite{oma_kompleksisuus_paperi}. We then discuss how
tighter bounds can be derived using n:th order density
correlations. In chaotic systems, the upper bounds considered here
provide a connection between the observables, assumed to follow the
ETH, and the arc shape of the complexity as a function of energy,
which we demonstrate numerically. We then study the behaviour of the
upper bounds over the localization transition, providing a connection
between the single-particle and Hilbert-space descriptions, and
discuss how this appears in quench dynamics.

\section{Upper bounds for Renyi-complexities}

\subsection{Formal setup and derivation of results}

In this section we will discuss a class of observable-based measures
that can be used to quantify the complexity of a quantum state. If we
fix a basis, the state can be expanded as
\begin{equation}
  \ket{\psi} = \sum_{n} a_n \ket{n},
\end{equation}
and the corresponding Renyi-$q$ complexities can be defined as the
entropies of the weight distribution in this basis,
\begin{equation}
  S_q(\vec{p})=\log \left( \sum_n p_n^q \right)/(1-q),
\end{equation}
with the Shannon limit defined as
\begin{equation}
  S_1(\vec{p})= - \sum_n p_n \log \left( p_n \right),
\end{equation}
where $p_n=|a_n|^2$. Suppose now that we have information on this
state in the form of expectation values of observables $\hat{d}_i$
diagonal in this basis,
\begin{equation}
  \braket{\psi | \hat{d}_i | \psi} = \sum_n p_n \braket{n|\hat{d}_i|n}  = d_i,
  \label{eqn:entropy_constraints_d}
\end{equation}
where the $d_i$ are known values. This knowledge limits the possible
complexity, and finding the upper limit correponds to maximizing the
entropy $S_q$ under the constraints given by Eqn.
\ref{eqn:entropy_constraints_d} and the additional requirement that
the weights form a probability distribution,
\begin{equation}
  \begin{split}
    & S_q^{max} = \max_{ \vec{p} } S_q(\vec{p}), \\
    & \text{    subject to    } \\
    & \sum_n p_n \braket{n|\hat{d}_i|n} = d_i,~i=1...N. \\
    & p_n \geq 0,~n=1...Q \\
    & \sum_n p_n = 1 \\
  \end{split}
  \label{eqn:general_optimization_problem}
\end{equation}
This is generally a convex optimization problem that can be solved
numerically \cite{boyd2004convex}, and we will present results for small enough systems
below. However, in the case of the Shannon entropy $S_1$ it is also
possible to find analytical results based on the similarity of this
problem to standard thermodynamics. In App.
\ref{app:analytical_results_derivation} we consider a fermionic
system, and assume that the average occupation of each of the orbitals
is known, i.e. that $\hat{d}_i$ are the occupation number
operators $\hat{d}_i=\hat{n}_i$. We denote the resulting maximum
complexity as $S_1^{max,n} \geq S_1$. As shown in the appendix,
$S_1^{max,n}$ then has an upper bound that is formally the same as the
grand canonical entropy of free fermions,
\begin{equation}
  \begin{split}
    S_1^{max,gc} &=-\sum_i \left( n_i \log(n_i) + (1-n_i) \log(1-n_i) \right) \\
     &\geq S_1^{max,n} \geq S_1,
  \end{split}
\end{equation}
where $n_i$ are the average occupations of the orbitals. The bound is
larger than the exact $S_1^{max,n}$ because it neglects particle
conservation constraints that are typically preserved by the
Hamiltonian, and which we tacitly assume to hold in
Eq. \ref{eqn:general_optimization_problem}. Noting the general
inequality $S_{q_1} \leq S_{q_2}$ for $q_2 < q_1$, we also see that
$S_1^{max,n}$ bounds all Renyi-complexities from above. While we only
consider fermions here, we also note that the discussed ideas are more
general and analogous bounds can be derived e.g. for bosonic or
spin-systems.

$S_1^{max,gc}$ yields immediately an upper bound also for the fractal
dimension $D_1$ as $D_1^{max,gc}=S_1^{max,gc}/\log(Q)$, where $Q$
is the number of basis states. Due to the neglected particle
conservation, this bound is generally not very tight for numerically
attainable system sizes, and can even yield values larger than
$1$. However, as shown in
App. \ref{app:analytical_results_derivation}, the thermodynamic limit
of the bound, which we call $D_1^{max,lim}$, can be expressed as
\begin{equation}
  \begin{split}
    D_1^{max,lim} &= \lim_{N_o \to \infty} D_1^{max,gc} \\
    &= \frac{ \lim_{N_o \to \infty} \frac{1}{N_o} S_1^{max,gc} }{ - \nu \log(\nu) - (1-\nu)\log(1-\nu)},
  \end{split}
  \label{eq:single_particle_D_thermodynamic_limit}
\end{equation}
where $N_o$ is the number of orbitals and $\nu$ is the overall filling
fraction of the system. Thus it is the intensive quantity
$S_1^{max,gc}/N_o$ that determines the upper bound for large systems,
always yielding $\lim_{N_o \to \infty} D_1^{max,gc} \leq 1$. Below we
use $D_1^{max,lim}$ also for finite systems with the understanding
that $\lim_{N_o \to \infty} S_1^{max,gc}/N_o$ is approximated from a
system with finite $N_o$. It is also important to note that
$S_1^{max,gc}-S_1^{max}$ is not expected to converge to zero even for
arbitrarily large systems due to the neglected particle conservation,
but instead may diverge. However, $D_1^{max,gc}$ and $D_1^{max}$ are
expected to converge to the same value, as the error grows slower than
$N_o$. In practice we find that $S_1^{max,gc}/N_o$ converges faster
than $S_1^{max,gc}/\log(Q)$, and thus
Eq. \ref{eq:single_particle_D_thermodynamic_limit} is preferred for
estimating the thermodynamic limit, as numerically demonstrated below.

One can also consider complexity bounds under constraints on higher
order correlation functions. For example, we can set
$\hat{d}_i=\hat{n}_{\mu_i} \hat{n}_{\nu_i}$, asking for the highest
possible complexity when the density-density correlators are known, or
$\hat{d}_i=\hat{n}_{\mu_i} \hat{n}_{\nu_i} \hat{n}_{\gamma_i}$ if the
third order correlators are known, and so on. We denote these higher
order bounds as $S_q^{max,nn}$, $S_q^{max,nnn}$, etc. As discussed in
App. \ref{app:analytical_results_derivation}, computing these bounds
from the correlation functions requires solving classical lattice gas
problems which do not have an analytical solution, and thus we cannot
hope for a similar simple formula as in the case of one-particle
constraints, but the bounds can still be computed numerically.

\subsection{Discussion and comparison to earlier results}

The expression for $S_1^{max,gc}$ can be compared with the
``occupation number entropy''
\begin{equation}
  S_{occ}=-\sum_i n_i \log(n_i),
\end{equation}
used in Ref. \cite{PhysRevLett.115.046603} to quantify the numerically
observed transition in the single-particle occupation spectrum of the
natural orbitals when crossing the many-body localization transition.
Qualitatively, $S_{occ}$ and $S_1^{max,gc}$ very similarly measure the
uniformity of the occupation distribution, but $S_1^{max,gc}$ is
particle-hole symmetric. The relation $S_1^{max,gc} \geq S_1$ gives
additional support to the idea that such single-particle quantities
can be used to characterize the many-body localization transition, as
it formally relates the single-particle and Fock-space pictures.

The natural orbitals, defined as eigenorbitals of the one-particle
reduced density matrix \cite{lowdin1}, have been studied as a basis
where the full state vector has a compact representation
\cite{Bell_1970,Borland_1970,10.1063/1.1671034,Giesbertz_2014,coe_paterson_2015,lowdin1,jotain}. As
briefly discussed in App. \ref{natural_orbitals_appendix}, both
$S_1^{max,gc}$ and $S_{occ}$ have the property that, among all
possible choices of single-particle orbital sets, they attain their
minimum value in the natural orbital basis. This is one sense in which
the natural orbital basis is optimal: It does not necessarily give the
lowest complexity $S_1$, but it has the lowest upper bound to the
complexity if only the one-particle reduced density matrix is known.

Another relevant single-particle quantity is the local polarization
\cite{PhysRevB.104.174201} of spin systems, or its fermionic analogue,
the ``local purity'' \cite{PhysRevB.103.214206},
\begin{equation}
  S_{lp}=\frac{1}{N_o} \sum_i (2 n_i - 1)^2,
\end{equation}
which can be expressed in terms of the expectation value of the
Hamming distance $x_{ij}$ between the basis states $i$ and $j$
\cite{PhysRevB.103.214206},
\begin{equation}
  S_{lp}=1-\frac{2}{N_o}\sum_{ij} x_{ij} p_i p_j.
\end{equation}
The local purity has been related to the second Renyi-complexity
$S_2$, but the exact relationship requires more information on the
state than the upper bound relation discussed above for the Shannon
entropy. In the many-body localized state the eigenstates typically
concentrate around one Fock configuration, and the weights decay
exponentially with increasing Hamming distance between the occupation
number sets \cite{Roy_2020}, which leads to a linear relationship
between $S_{lp}$ and the inverse participation ratio $\exp(S_2)$
\cite{PhysRevB.104.174201}. Numerically a relationship between
$S_{lp}$ and related quantities and $S_2$ is observed also e.g. for
eigenstates of many-body Hamiltonians and random matrix ensembles
\cite{oma_kompleksisuus_paperi}. It is thus an interesting question
whether the solution to Eq.  \ref{eqn:general_optimization_problem}
with $q=2$ can be expressed in terms of the local purity similarly to
how the $S_1^{max,gc}$ provides the limit for the $q=1$ case. However,
we are unaware of any useful analytical solutions for the optimization
problem Eqn. \ref{eqn:general_optimization_problem} for $q>1$.

\section{Numerical results}

The problem \ref{eqn:general_optimization_problem} is a convex
optimization problem whose global optimum can be found using numerical
methods \cite{boyd2004convex}. Here we have employed the QuSpin
library for building the Hamiltonian matrices and for finding the
eigenstates
\cite{10.21468/SciPostPhys.2.1.003,10.21468/SciPostPhys.7.2.020}, and
the CVXOPT-library \cite{cvxopt_library} for solving the optimization
problem. We perform the optimization either with single-particle
conditions, where the density expectation values are assumed to be
known, and with two-particle conditions where density-density
correlation functions between all orbitals are given. We also present
some results with three-particle conditions, where all
density-density-density correlations are known, but naturally the
optimization problem becomes more difficult to solve with increasingly
complex conditions. We note that, because of the particle
conservation, the $n$-particle conditions imply the $n-1$ particle
conditions, and thus always result in a stricter bound. In practice we
first solve a selected eigenstate of the system, which allows us to
compute the weights $p_i$, the exact $S_q$, the occupations
$\left\langle n_i \right\rangle$, and the higher correlation
functions, and then proceed to the optimization. We discuss results
for the Shannon entropy $S_1$ and the Renyi-$2$ entropy $S_2$.

We demonstrate the upper bounds for the $t-V$-model defined as
\begin{equation}
\begin{split}
  H &= \sum_{\langle i, j\rangle} \Bigg( -t \left( \hat{c}^\dagger_{i}\hat{c}_{j}+ \mathrm{h.c.} \right) \\
  &+ V \left( \hat{n}_{i}-\frac{1}{2} \right) \left( \hat{n}_{j}-\frac{1}{2} \right) + \epsilon_i \left(n_i - \frac{1}{2} \right) \Bigg), \\
\end{split}
\end{equation}
where we set $t=\frac{1}{2}$ and $V=1$. The random on-site potential
$\epsilon_i$ is taken to be uniformly distributed in the interval
$[-W,W]$. For small enough $W \lesssim 3$ the model is in a chaotic,
thermalizing phase. When $W$ is increased, the model transitions into
the many-body localized region, although the exact nature and location
of the localization transition is still under debate
\cite{sierant2024_mbl_review}. Here we will first discuss the
arc-shaped form of the fractal dimensions as a function of the energy
in the chaotic phase, and then study the localization transition.

\subsection{Chaotic phase}

\begin{figure}
  \includegraphics[width=\columnwidth]{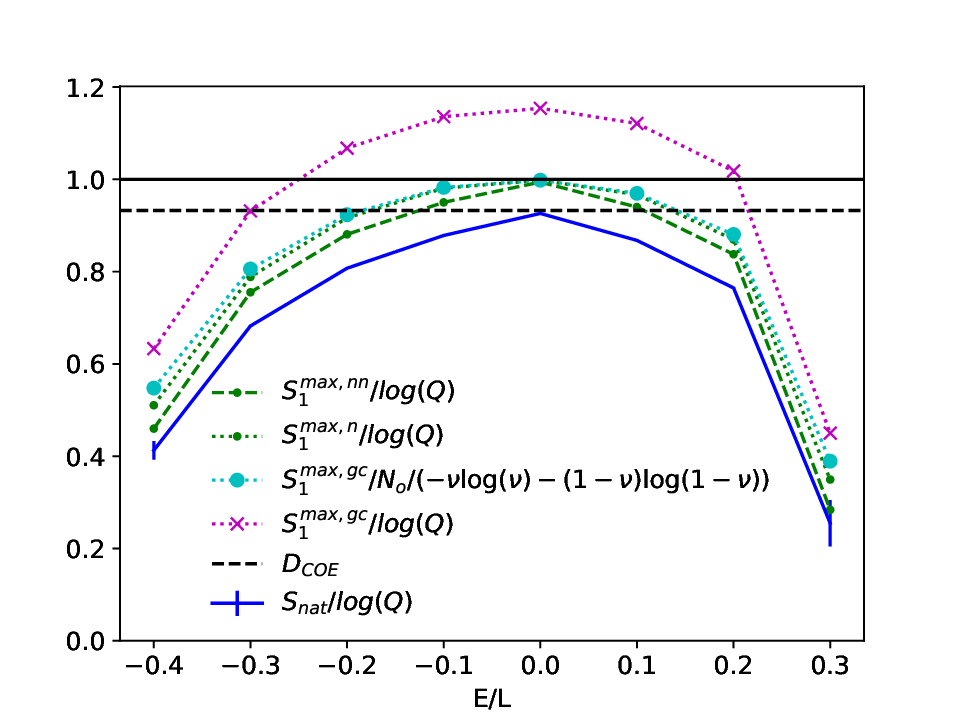}
  \caption{Fractal dimension $D_1$ and different upper bound estimates
    for the $t$-$V$ -model in the natural orbital basis. The model is
    in the chaotic phase with $t=W=V=1.0$ and the chain length is
    $L=18$. The results are computed as the mean over $1000$ random
    potential realizations at each energy, with the (small) error bars
    on $D_1=S_{nat}/\log(Q)$ showing the standard error. The
    horizontal lines demarcate the trivial bound $D_1=1$ and the
    result for the circular orthogonal ensemble.}
  \label{fig:tV_model_chaotic_phase_S_1}
\end{figure}

\begin{figure}
  \includegraphics[width=\columnwidth]{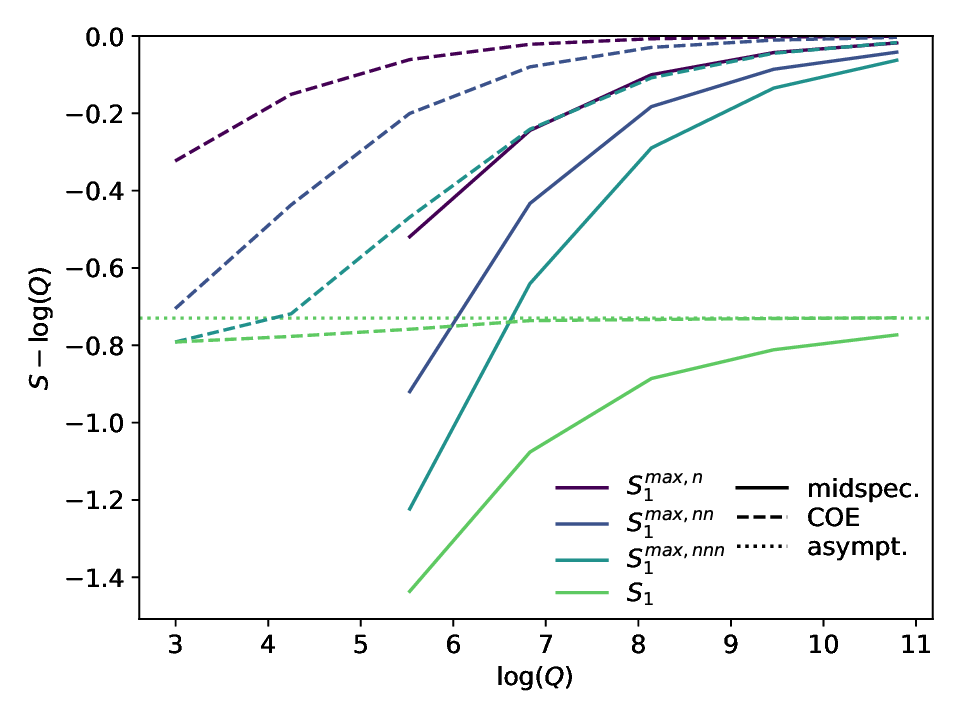}
  \caption{The complexity $S_1$ and its $1$, $2$ and $3$-body upper
    bounds as a function of the system size for the COE and the $t -
    V$ model at midspectrum. The results are computed in the natural
    orbitals, and averaged over $1000$ disorder realizations. For each
    disorder realization of the $t-V$ model we compute the mean energy
    in the canonical ensemble at infinite temperature, and then select
    an eigenstate close to this energy. According to the ETH, this is
    expected to yield states that have no few-body structure in the
    thermodynamic limit. We have subtracted the trivial contribution
    $\log(Q)$ from all plotted quantities to better show the
    differences. For the COE, $S_1$ approaches the asymptotic result
    $S_1=\log(Q) - c$, with $c$ a known
    constant.\label{fig:midspectrum_S_scaling}}
\end{figure}

Fig. \ref{fig:tV_model_chaotic_phase_S_1} shows the fractal dimension
$D_1$ and the associated one- and two-particle bounds for the $t - V$
model in the chaotic phase. The fractal dimension shows the typical
arc shape, with an apex at $E/L \approx 0$ in agreement with the fact
that the expected energy per length unit of the system approaches zero
at infinite temperature in the large system limit. The loosest bound
is the grand canonical approximation to the single-particle bound,
which, at this system size at midspectrum, is well above the trivial
bound $D_1 \leq 1$. However, the thermodynamical limit formula, Eq.
\ref{eq:single_particle_D_thermodynamic_limit}, closely follows the
numerically computed single-particle bound that takes into account
particle conservation. This shows that, already at this small system
size, Eq. \ref{eq:single_particle_D_thermodynamic_limit} provides a
useful upper bound estimate. The two-particle bound is somewhat
tighter than the one-particle bound, but the improvement in this case
is not drastic. This is understandable, because the system with $V=1$
is weakly correlated. In the limit of small $V$ we may expect that
Hartree-Fock mean-field theory gives a good description of the
correlation functions, which implies that the two-particle
correlations factorize and thus give no additional information on the
state.

It is interesting to discuss how tight the bounds considered here can
become. If we consider generic $n$-particle conditions on the weight
distribution, the bound becomes tighter with increasing $n$, giving
the exact $S_q$ when $n=N_p$. However, the interesting limit is rather
the case where $n$ is small and fixed and the system size approaches
infinity, i.e. the ``few-body'' limit. According to the eigenstate
thermalization hypothesis, the few-body observables in large systems
should be predicted by a thermal (microcanonical or canonical)
ensemble \cite{Polkovnikov2016ETH,Deutsch_2018,PhysRevE.50.888}, and
thus the upper bounds only depend on the energy or temperature. At
``midspectrum'', where the temperature is infinite and observables
show no structure, all such few-body bounds should approach the
trivial bound $S_{q}^{max}=\log(Q)$. However, the actual complexity,
$S_q$, does not approach the trivial bound as the state generally
contains structure not seen in few-body observables. A common
assumption is that the structure can be described by the eigenstates
of a random Hamiltonian ensemble that respects symmetries of the
problem, such as the gaussian orthogonal ensemble (GOE) when the only
relevant symmetry is time-reversal
\cite{Polkovnikov2016ETH,PhysRevE.100.032117}. For the GOE eigenstates
(distributed according to the circular orthogonal ensemble (COE)), the
large-$Q$ limit is given by $S_1 \approx 1 - c$, where $c \approx
\left(\log(2) + \psi(3/2) \right)$ and $\psi$ is the digamma function
\cite{PhysRevE.100.032117}. Thus we may expect that a gap,
representing ``random fluctuations'' not captured by few-body
observables, remains between $S_q$ and the $n$-body bounds regardless
of $n$, as long as $n \ll N_p$. The question is if this gap generally
remains bounded, as expected at midspectrum, or if it grows linearly,
leading to a constant gap in the fractal dimension in the
thermodynamic limit. We will consider this question below, but we
first present numerical results at midspectrum.

\begin{figure}
  \includegraphics[width=\columnwidth]{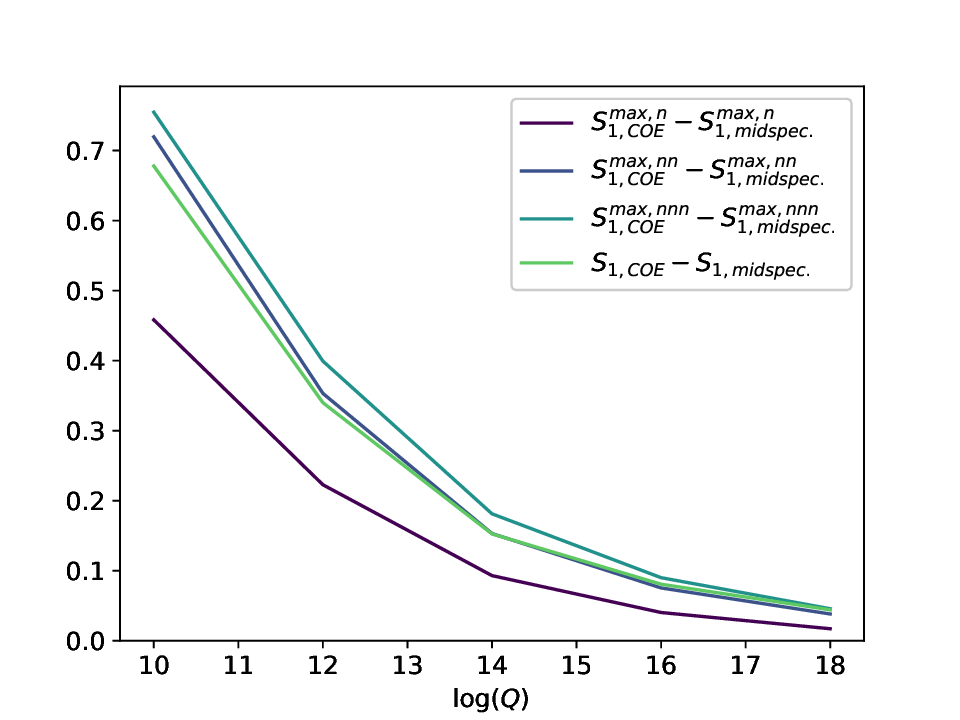}
  \caption{Difference between the corresponding COE and $t-V$ model
    quantities of
    Fig. \ref{fig:midspectrum_S_scaling}.\label{fig:midspectrum_S_diff_scaling}}
\end{figure}

To test the above picture at midspectrum, we compare the upper bounds
and complexities of the $t - V$ model to those of COE-distributed
states in Fig. \ref{fig:midspectrum_S_scaling}. As has been noted for
other models in the literature \cite{PhysRevE.100.032117}, the
finite-size corrections to $S_1$, or to the fractal dimensions,
generally differ from the random matrix predictions. This is observed
also here, as the complexity of the COE states converges much faster
to the asymptotic result $S_1 \approx \log(Q) - c$ compared to the $t
- V$ model. Interestingly, the same behaviour is seen also in the
upper bounds: The $n$-particle bound for the COE approaches $\log(Q)$
faster than for the $t-V$ model. Thus we can conclude that the slower
approach to the thermodynamic limit is due to the few-body
correlations present in the $t-V$ eigenstates, and the upper bounds
provide an economical measure of such correlations. Such structure is
expected, since the $t-V$ Hamiltonian, and indeed most physical
Hamiltonians, are composed of few-body terms. This can be compared to
earlier results on the XYZ spin chain showing that residual spatial
correlations can reduce the midspectrum entanglement entropy in finite
systems \cite{PhysRevE.105.014109}.

\begin{figure*}[t]
  \includegraphics[width=\textwidth]{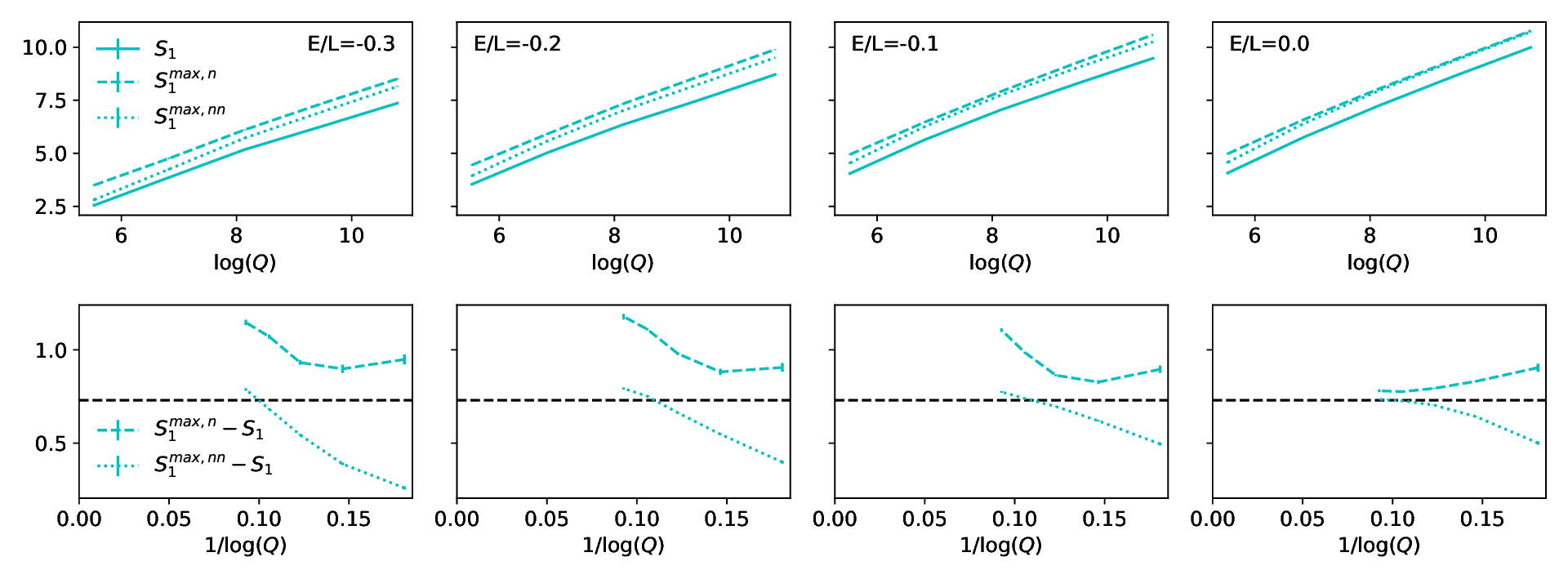}
  \caption{Upper panels: Complexity $S_1$ and its one- and
    two-particle upper bounds as a function of the logarithm of the
    Hilbert space dimension in natural orbital basis. Lower panels:
    The gap between the actual complexity and the upper bounds. The
    dashed black line is the asymptotic COE gap $c$. The results have
    been averaged over $100$ disorder realizations away from $E=0$ and
    $1100$ realizations at $E=0$, except for the largest system size
    $L=18$ where we used only $10$ realizations in all cases. The
    standard error is indicated where it is larger than the line
    width.
    \label{fig:away_from_midspectrum_gap}}
\end{figure*}

We also plot the difference of the upper bounds between the COE states
and the $t-V$ model in Fig. \ref{fig:midspectrum_S_diff_scaling}, thus
measuring the excess of $n$-body structure in the $t-V$ model
eigenstates compared to the COE states. Notably, this excess for the
two- and three-body bounds roughly corresponds to the difference of
the actual complexities between the two systems. This supports the
idea that the complexity can be thought of as a sum of two terms: A
term arising from a few-body structure, which is smaller for the $t-V$
model compared to the COE states, and a random term which is similar
for the two models.

Moving away from the infinite temperature point, we compare the
scaling of the upper bounds and the true complexity at fixed specific
energies. Here the most interesting question is if the gap between the
true complexity and the upper bounds saturates to a finite value, or
if it tends to infinity in the large system size limit. The complexity
and the upper bounds are plotted in
Fig.\ref{fig:away_from_midspectrum_gap}. The gap generally increases
with system size, but, plotted against $1/\log(Q)$, it does not appear
divergent, especially considering the two-particle bound. Thus it
seems plausible that the few-body bounds can correctly predict the
fractal dimension of the state in the infinite system size limit, with
a constant gap remaining between the bound and the
complexity. Intuitively, the few-particle structure limits the Hilbert
space volume available to the state to a fraction of the total volume
$Q$, and this fraction vanishes in the thermodynamic limit, except for
the infinite temperature states. The states, however, fill some finite
fraction of that limited volume, which is reflected in the size of the
gap. This remaining gap represents a ``randomness'' of the state that
cannot be captured by few-body observables. This picture summarizes
our ``hypothesis of the arc'' that relates the complexity in chaotic
systems to few-particle observables predicted by the ETH.

\subsection{Localization transition}

\begin{figure*}[t]
  \includegraphics[width=\textwidth]{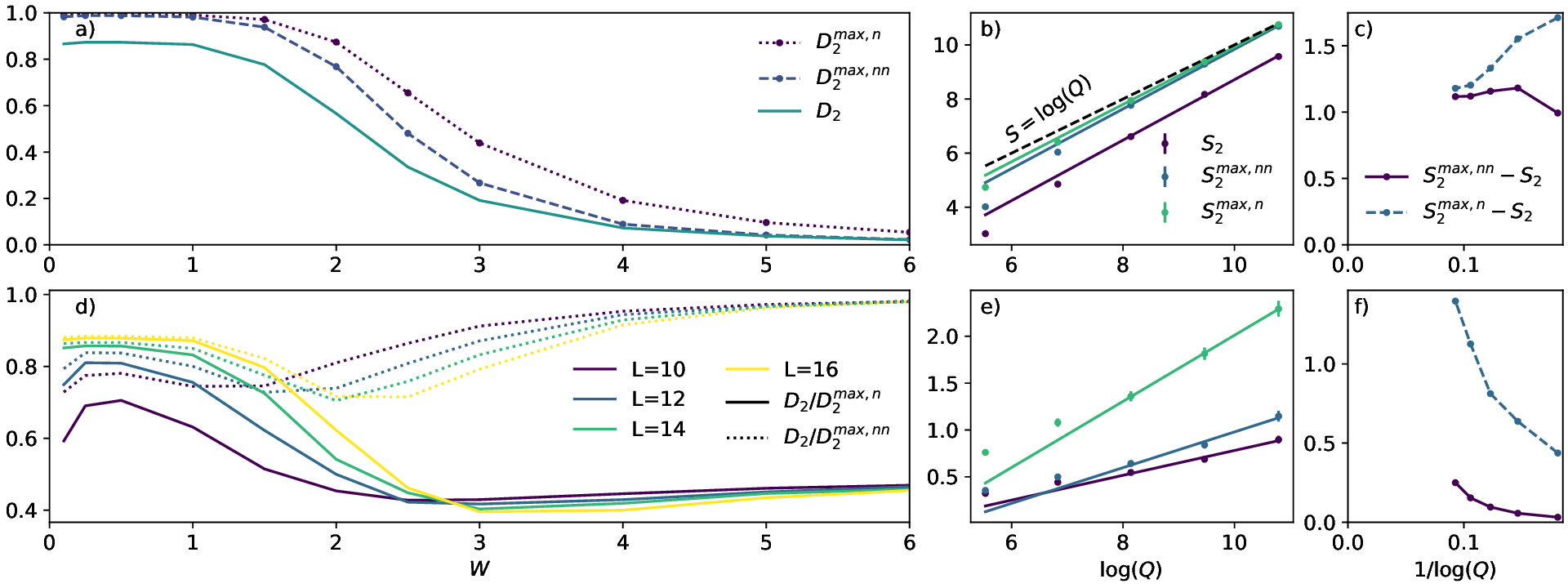}
  \caption{Complexity upper bounds over the localization transition in
    the natural orbital basis. a) The midspectrum ($E/L=0$) fractal
    dimension $D_2$ and its upper bounds for a system of size $L=16$
    computed as a mean over $1000$ disorder realizations.  b) System
    size dependence of the Renyi-$2$ complexity $S_2$ and its upper
    limits at $W=1$ (chaotic system). Solid lines are linear fits to
    the $3$ rightmost points.  c) The gap between the bounds and the
    actual complexity $S_2$ as a function of $1/\log(Q)$ similarly as
    in Fig. \ref{fig:away_from_midspectrum_gap}.  d) The mean ratio of
    the true fractal dimension to the upper bounds for different
    systemn sizes.  e-f) As in b-c but for $W=4$ (localized region).
    \label{fig:localization_transition}}
\end{figure*}

The existence, nature and location of many-body localization
transitions in the thermodynamic limit are still not fully understood
\cite{sierant2024_mbl_review}. In finite systems accessible by exact
diagonalization methods the transition is clearly signalled by the
level spectrum, the eigenstate fractal dimensions and single-particle
observables
\cite{sierant2024_mbl_review,PhysRevLett.115.046603}. Scaling analyses
indicate that the fractal dimension undergoes a non-universal jump at
the transition point \cite{PhysRevLett.123.180601}, thus sharply
defining the transition. However, due to the system size limitations,
the correct scaling ansatz has not been definitively identified, and
it is typical that the transition point seems to ``creep'' to higher
disorder strenghts with sufficiently large system sizes
\cite{sierant2024_mbl_review}. Thus the possibility remains that this
is not a true phase transition, but a crossover that moves to infinite
disorder strenghts in the thermodynamic limit. Here we do not attempt
to solve this long-standing puzzle, but to discuss how the localized
states at attainable system sizes are characterized by the upper
bounds.

Fig.~\ref{fig:localization_transition}a shows the fractal dimension
$D_2$ and its upper bounds at midspectrum as the disorder strength is
increased from the chaotic region. We discuss here the Renyi-$2$ case
as has been previously considered in the context of MBL
\cite{PhysRevB.103.214206,PhysRevB.104.174201} and because we find the
optimization problem, Eqn. \ref{eqn:general_optimization_problem},
numerically harder to solve for the $q=1$ case in the localized
region. The behaviour in the chaotic region at midspectrum parallels
that of the Renyi-$1$ case with the upper bounds approaching
$S=\log(Q)$, and the complexity $S_2$ retaining a gap below the bounds
that approaches a finite value, as shown in
Fig. \ref{fig:localization_transition}b and
\ref{fig:localization_transition}c. The fractal dimension
$D_2=S_2/\log(Q)$ thus converges to $D_2 \approx 1$ in the chaotic
region as expected.

In contrast to the chaotic region, the localized region shows a clear
decrease in the estimated fractal dimension and the upper bounds,
resembling the occupation number entropy of reference
\cite{PhysRevLett.115.046603}. Another difference to the chaotic case
is that the fractal dimension predicted by the upper bounds is now
clearly higher than the true fractal dimension in the large system
limit. This is seen e.g. in Fig. \ref{fig:localization_transition}d
where the ratio of $D_2$ to its upper bounds is plotted. In the
chaotic region these ratios increase with increasing system size,
while for large enough disorder strength they start to decrease. In
other words, the bounds $S_2^{max,n}$ and $S_2^{max,nn}$ in the
localized region grow faster than the true $S_2$, and the gap between
$S_2$ and the bounds increases proportionally to $\log(Q)$, as seen in
Fig.~\ref{fig:localization_transition}e and
\ref{fig:localization_transition}f. This can be contrasted to the
results in the chaotic phase shown in
Fig.~\ref{fig:away_from_midspectrum_gap}, where the increase is slower
and the gap appears to approach a finite value in the infinite system
size limit.

The natural orbitals can be seen as a single-particle approximation to
the emergent integrals of motion characterizing the localized states,
especially deep in the localized region \cite{2017AnP...52900356B}. In
this region the natural occupation spectrum resembles a fermi-liquid
state with one highest-weight Slater configuration and a discontinuity
at the ``fermi surface'' \cite{2017AnP...52900356B}. In contrast, the
occupation spectrum in the chaotic region resembles a
finite-temperature fermi distribution and the discontinuity is
expected to vanish in the thermodynamic limit. Consistently with this
picture, the one-particle bound, which is related to the ensemble of
free fermions (see App. \ref{app:analytical_results_derivation}), is
relatively close to the two-particle bound in the chaotic region,
while the two-particle conditions give a significant improvement in
the localized region. We can also say that the two-particle
correlations, resulting from interaction effects, are much more
important in the localized region.

\begin{figure}[h]
  \includegraphics[width=\columnwidth]{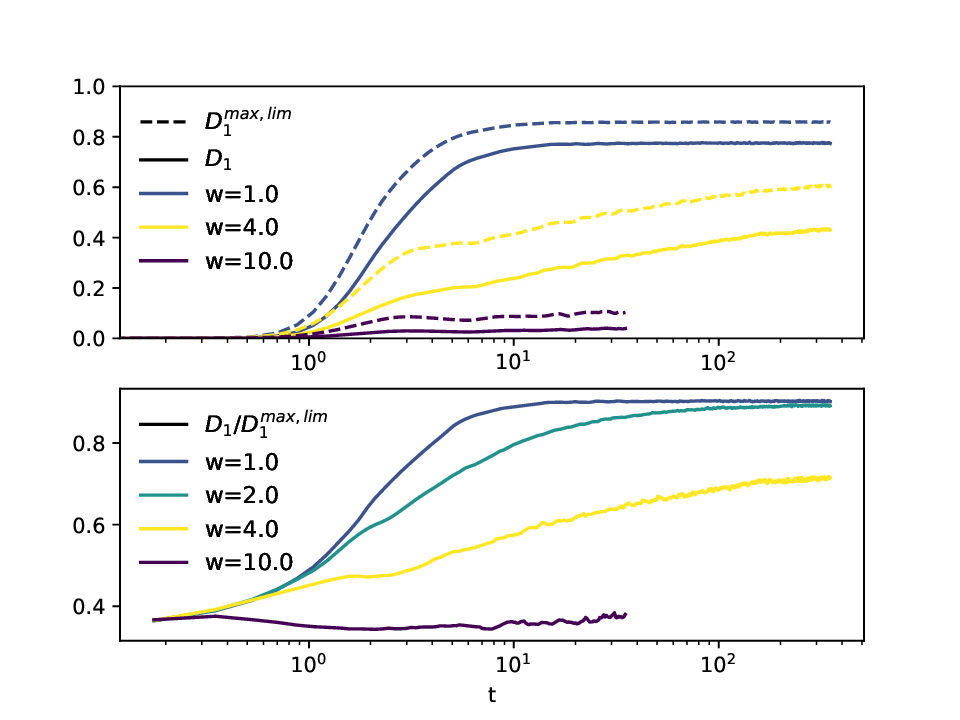}
  \caption{Upper panel: Time development of the mean fractal dimension
    and the thermodynamical limit formula, Eq.
    \ref{eq:single_particle_D_thermodynamic_limit} in the natural
    orbital basis. The system starts from the Slater state with
    alternating occupations $\ket{010101 \dots}$ and is time developed
    with the $t - V$ Hamiltonian of length $L=14$ with varying
    disorder strength. The results are averaged over $100$ disorder
    realizations. The limit $\lim_{N_o \rightarrow \infty} S_1/N_o$ is
    estimated directly from the data at this system size. Lower panel:
    ratio of the complexity and the limit
    formula. \label{fig:dynamics}}
\end{figure}

We finally consider the behaviour of the fractal dimension and the
upper bounds for quenches starting from a Slater determinant
state. Such quenches have been used in numerical experiments e.g. to
investigate the relationship between dynamics of the entanglement
entropy and physical observables \cite{PhysRevB.108.134204}. Here we
want to see if the one-body bound $D_1^{max,lim}$ acts as a proxy
quantity to the fractal dimension $D_1$ as it does for the
eigenstates. Indeed, as shown in Fig. \ref{fig:dynamics}, these
quantities qualitatively follow each other, starting from zero in a
Slater configuration and developing into a plateau with slowly
increasing complexity. This slow increase may be another symptom of
the ``creeping'' phenomenon where the system slowly seems to develop
towards the chaotic system (large complexity) even for large disorder
strengths, where the complexity initially seems to saturate to lower
values.

\section{Summary and outlook}

In this work we have discussed how structure measured in few-body
observables limits the complexity and fractal dimensions of
eigenstates of many-body systems. By analogy to a thermal free-fermion
system, we derived an upper bound for the Shannon-type fractal
dimension $D_1 \leq D_1^{max,lim}$
(Eq.~\ref{eq:single_particle_D_thermodynamic_limit}), which requires
the knowledge of the occupation numbers in the chosen orbital
basis. In our chaotic model system the formula predicts the arc shape
of $D_1$ as a function of energy quite well. The bound can be improved
if information on higher order correlation functions, such as
density-density correlations, is available, which leads to a series of
bounds with increasing correlator order. For weakly coupled chaotic
systems, where few-body correlators are well described by thermal
mean-field theory, the higher order correlators factorize and
therefore give no additional information, and thus the lowest order
bound $D_1$ is expected to be quite tight. This argument is based on
the ETH, i.e. that few-body observables are assumed to follow thermal
averages, and one motivation for our study is to connect the thermally
predicted observables to the complexity.

We also considered how the upper bounds can be used to locate and
characterize the many-body localization transition. In earlier
literature the transition has been characterized using both
single-particle quantities and fractal dimensions
\cite{PhysRevLett.115.046603,2017AnP...52900356B,PhysRevB.103.214206,Polkovnikov2016ETH,Borgonovi_2016,PhysRevB.103.214206,PhysRevLett.123.180601,PhysRevE.98.022204,De_Luca_2013,Santos_2010,Beugeling_2015,PhysRevB.91.081103,Misguich_2016,Tsukerman_2017,PhysRevLett.124.200602}. Our
work formally connects these two pictures, as deviation of e.g. the
one-body upper bound from the maximal value immediately implies
non-ergodicity of the state (i.e. fractal dimension $D_1 <
1$). Furthermore, the higher order bounds act as a measure of
correlation, separating the localized states from mean-field -like,
weakly correlated states. We also showed that the one-body bound acts
as a more easily studied proxy-quantity to the complexity and fractal
dimension in a dynamical setting, which could be useful e.g. in
ultracold gas experiments.

To further understand behaviour of the derived bounds in the
thermodynamic limit, we analyzed their system size scaling first in
the chaotic model. At midspectrum, where few-body observables are
given by the infinite-temperature ensemble, all few-body bounds
$S_1^{max}$ approach $\log(Q)$ in the thermodynamic limit and the gap
$S_1^{max}-S_1 \rightarrow c$, where $c$ is well predicted by a
suitable random matrix ensemble respecting the symmetries of the
system. However, as noted in the literature, the finite-size scaling
of $S_q$ and $D_q$ are system dependent and deviate from the COE
predictions \cite{PhysRevE.100.032117}. Here we employed the $1$-
,$2$- and $3$-body bounds to show that our model system has an excess
of few-body correlations compared to the COE, explaining the slower
approach to the thermodynamic limit. The bounds thus act as a measure
of correlations that are useful for characterizing deviations from
random matrix predictions.

When moving away from midspectrum to finite temperature states, it
still seems that the gap between the actual complexity and the
few-body bounds remains bounded in the thermodynamic limit,
$S_1^{max,nn}-S_1 \rightarrow C$ for example. This also means that the
fractal dimension $D_1$ in the thermodynamic limit is exactly
predicted by the corresponding bound $S_1^{max,nn}/\log(Q)$. While the
bound $S_1^{max,nn}$ represents a limitation to the ``volume'' of the
state set by the thermal correlations, the gap $C$ includes the
contribution of ``random fluctuations'' not visible in any few-body
observables, and remains similar in magnitude to the gap at
midspectrum. An interesting interpretation is thus that eigenstates in
a chaotic system are still ``random states'' even away from
midspectrum, but now within the limitations of the few-body
observables prescribed by the ETH, although the precise definition of
the random ensemble in question remains a subject for future work. The
above picture is different in the localized region of our model, where
it seems that the gap between the few-body bounds and the complexity
grows linearly with $\log(Q)$ and thus the fractal dimension is not
exactly given by the respective upper bounds. The intuitive picture is
thus that the ``volume'' of the state grows slower than expected from
the bounds.

We note that the conclusions on the scaling behaviour of the bounds
are based on exact diagonalization calculations with limited system
sizes, and further results from different model systems would be
desirable to confirm the picture. In the chaotic phase, under
assumption of the ETH, it may be possible to study the bounds via
e.g. Monte-Carlo calculations, as they only require knowledge of
selected observables instead of the full state vector. In such
calculations the complexity itself remains unknown, but calculating
successive $n$-body bounds might reveal interesting information.

In summary, we have considered a new class of upper bounds to the
complexity that are useful for understanding the arc-shape of the
fractal dimension as a function of energy in chaotic systems, can be
employed as a measure of correlations to analyze deviations from
random matrix predictions, and to formally connect the
observable-based and fractal-dimension-based pictures of many-body
localization. We thus expect these quantities to give new insights in
the study of quantum chaos and many-body localization. An interesting
idea for future development would be to try to formulate a refined
random matrix model where the one-body or few-body bounds could be
tuned by parameters to model states away from midspectrum in a chaotic
system. As another future direction it would be interesting to
consider the complexity and its upper bounds as internal time scales
of quench dynamics similarly to the entanglement entropy in
\cite{PhysRevB.108.134204}.

\begin{acknowledgments}
The authors acknowledge the Academy of
Finland project 331094 for support. Computing resources were provided
by CSC -- the Finnish IT Center for Science.
\end{acknowledgments}

\bibliography{upperbound}

\appendix

\section{Derivation of analytical upper limit results}

\label{app:analytical_results_derivation}

Specializing to the case $q=1$, we will now seek solutions to the
optimization problem of
Eqn. \ref{eqn:general_optimization_problem}. Analogously to the
derivation of classical statistical ensembles, we can solve the
constrained maximization problem using lagrange multipliers. To this
end, we write a lagrangian
\begin{equation}
\begin{split}
  L(\vec{p},\vec{\mu}) &= \vec{p} \cdot \log(\vec{p}) - \sum_i \mu_i \left( \sum_{j} p_{j} \lambda_{ij} - \lambda_i \right) \\
  & - M \left( \sum_j p_j - 1\right),
\end{split}
\end{equation}
where the lagrange multipliers $\mu_i$ correspond to the expectation
value conditions and $M$ enforces the normalization. Taking
derivatives with respect to $p_i$ and the lagrange multipliers we get
the equations
\begin{equation}
  \begin{split}
    & \log\left(p_{j}\right)+1-\sum_i \mu_i \lambda_{ij} - M = 0 \\
    & \sum_{j} p_{j} \lambda_{ij} = \lambda_i \\
    & \sum_j p_j = 1
  \end{split},
\end{equation}
so that
\begin{equation}
  p_j=\exp\left( \sum_i \mu_i \lambda_{ij} + M - 1 \right)=\exp\left( \sum_i \mu_i \lambda_{ij}\right)/Z,
\end{equation}
where the normalization has been absorbed to $Z$. We note that the
positivity conditions are automatically satisfied.

Further results require some concrete operators $\lambda_{i}$. A
simple example is to take a fermionic many-body system with $N$
orbitals and let $\hat{\lambda}_i$ be the occupation of the $i$:th
orbital, $\hat{\lambda}_i=\hat{n}_i$. The basis states are Slater
determinants that can be indexed by their occupation number sets
$\{n_i\}$ with each $n_i=0,1$. Thus the weights of the maximum
complexity configuration take the form
\begin{equation}
  p_{\{n_i\}} = \exp\left( \sum_i \mu_i n_i\right)/Z,
\end{equation}
which is of the same form as in an ensemble of free fermions with
$\mu_i$ the normalized single-particle ``energies''. If we allow
configurations that mix different particle numbers, the weights of the
configurations are exactly those of a grand canonical ensemble of free
fermions. Thus the maximal complexity is given by the well-known
expression of the grand canonical entropy,
\begin{equation}
  S_1^{max,gc}=-\sum_i \left( n_i \log(n_i) + (1-n_i) \log(1-n_i) \right).
\end{equation}
This upper limit is tight in the sense that the configuration
attaining the maximal complexity with the given occupations is
constructed. However, this only holds in the ``grand canonical''
sense, when different particle numbers are mixed. As many-body
Hamiltonians usually conserve the particle number, the bound could be
made more strict by considering the canonical ensemble where only
configurations with a fixed particle number are allowed. However, a
simple formula for the entropy of the canonical ensemble for free
fermions does not exist. In fact, when particle conservation is
enforced, the constraints cannot even be satisfied if the occupation
number set lies outside the polygon defined by the generalized Pauli
constraints \cite{Altunbulak_2008,reuvers2021generalized}, which
indicates that the exact solution for the particle conserving case is
complicated. Nevertheless, the grand canonical formula is still an
upper bound for the complexity even in the particle conserving case,
as it simply relaxes a constraint.

The upper bound relation $S_1 \leq S_1^{max,gc}$ immediately also
gives an upper bound for the fractal dimension $D_1=S_1/\log(Q)$,
where $Q$ is the number of basis states in the Hilbert space. For a
single-component fermion system with $N_o$ orbitals the full space has
$Q=\binom{N_o}{\nu N_o}$ states, where $\nu$ is the filling fraction,
and the maximal possible complexity, obtained for the uniform
distribution, is simply $\log(Q)$. From the general inequality
\begin{equation}
  \frac{1}{n+1} \exp(n H(k/n)) \leq \binom{n}{k} \leq \exp(n H(k/n)),
\end{equation}
where $H(p)=-p\log(p)-(1-p)\log(1-p)$, we find that
\begin{equation}
  \begin{split}
    \log(Q) &= -N_o \left( \nu \log(\nu) + (1-\nu)\log(1-\nu) \right)
    \\ &+ O(\log(N_o)),
  \end{split}
\end{equation}
and thus, in the limit of $N_o \rightarrow \infty$, the bound for the
fractal dimension becomes
\begin{equation}
  \lim_{N_o \to \infty} D_1^{max,gc} = \frac{ \lim_{N_o \to \infty} S_1^{max,gc}/N_o }{ - \nu \log(\nu) - (1-\nu)\log(1-\nu)},
\end{equation}
with the assumption that $S_1^{max,gc}/N_o$ converges as the
system size is increased. The maximal value of $S_1^{max,gc}/N_o$
occurs with the uniform configuration where each $n_i=\nu$, and equals
$-\nu \log(\nu) - (1-\nu)\log(1-\nu)$. This shows that, despite the
relaxation of particle conservation, $D_1^{max,gc} \leq 1$ provides a
non-trivial upper bound in the thermodynamic limit. In particular,
midspectrum states of chaotic systems are expected to have $D_1
\rightarrow 1$ in the thermodynamic limit, which implies that
$D_1^{max,gc}$ must approach $1$.

It is also possible to obtain some analytical results for conditions
involving higher correlators. For example, if we assume knowledge of
some subset of the density-density correlators $\braket{ \psi |
  \hat{n}_i \hat{n}_j | \psi}$, $ij \in S$, we get analogously to the
above derivation
\begin{equation}
  p_{\{n_i\}} = \exp\left( \sum_{ij \in S} \mu_{ij} n_i n_j\right)/Z,
\end{equation}
where the summation is over the chosen subset $S$ of pairs $i,j$. To
find the upper bound we should then determine the lagrange multipliers
$\mu_{ij}$ such that the expectation value constraints are
fulfilled. The general case corresponds to solving a lattice gas
problem or a classical Ising model with arbitrary coupling constants,
and no analytical solution is thus available. However, we can consider
e.g. the case of a 1D model in the position basis, where the
well-known solution of the Ising model can be applied. If only
knowledge of the average density-density correlation between
nearest-neighbour orbitals is assumed, and particle conservation is
again relaxed, then the result corresponds to the entropy of the 1D
Ising model expressed as a function of the nearest neighbour
correlation function.

\section{Notes on the natural orbitals}

\label{natural_orbitals_appendix}

The natural orbitals of a pure state $\ket{\psi}$ are defined as the
eigenorbitals of the one-particle density matrix
\begin{equation}
  \rho_{ij} = \braket{\psi|c_i^\dagger c_j|\psi},
\end{equation}
and the corresponding eigenvalues, i.e. the average occupations of the
natural orbitals, are referred to as the natural occupations. Let us
denote the occupations in some arbitrary orbital basis as $n_i$ and
the natural occupations as $\lambda_i$. We will assume that they are
indexed in decreasing order such that $n_i \geq n_{i+1}$ and
$\lambda_i \geq \lambda_{i+1}$. By the Schur-Horn theorem the natural
occupations majorize \cite{marshall2010inequalities} the occupations
in any other orbital basis \cite{coe_paterson_2015}, meaning
\begin{equation}
  \sum_{i=1}^k \lambda_i \geq \sum_{i=1}^k n_i,
\end{equation}
for all $k=1...N_o$, with
\begin{equation}
  \sum_{i=1}^k \lambda_i = \sum_{i=1}^k n_i.
\end{equation}
We also write in vector notation $\vec{\lambda} \succ \vec{n}$. If $S$
is any Schur-convex function, such as the Shannon entropy or any of
the Renyi-entropies $S_q$ with $q>0$, then $S(\vec{\lambda}) \leq
S(\vec{n})$ \cite{marshall2010inequalities}. It thus follows that
e.g. the occupation entropy $S_{occ}=-\sum_i n_i \log(n_i)$ is
minimized in the natural orbital basis. Furthermore, if we consider
the hole occupations ordered from largest to smallest,
$n_i^h=1-n_{N_o-i+1}$ and $\lambda_i^h=1-\lambda_{N_o-i+1}$, it can be
shown that the $\lambda^h_i$ majorize $n^h_i$, or $\vec{\lambda}^h
\succ \vec{n}^h$. It then follows that the hole occupation entropy
$S_{occ,h}=-\sum_i (1-n_i)\log(1-n_i)$ and the upper bound
$S_1^{max,gc}=S_{occ}+S_{occ,h}$ are also minimized in the natural
orbital basis. Therefore the smallest upper bound for the complexity
is obtained in the natural orbital basis.

\end{document}